\documentclass{llncs}
\usepackage{llncsdoc}

\usepackage{hyperref}
\usepackage{graphicx}
\usepackage{amsmath}
\usepackage{amssymb}
\usepackage{latexsym}

\usepackage[usenames]{color}
\usepackage[normalem]{ulem} %for strikethrough
\definecolor{MyRed}{rgb}{0.55,0,0}
\definecolor{MyGreen}{rgb}{0.0,0.55,0.0}

\begin{document}

\title{The Emergence of Norms via Contextual Agreements in Open Societies}

\author{George A. Vouros}
\institute{Digital Systems, University of Piraeus, Greece \email{georgev@unipi.gr}
}

\maketitle

\begin{abstract}
This paper explores the emergence of norms in agents' societies  when agents play multiple -even incompatible- roles in their social contexts simultaneously, and have limited interaction ranges. Specifically, this article proposes two reinforcement learning methods for agents to compute agreements on strategies for using common resources to perform joint tasks.  The  computation of  norms by considering agents' playing multiple roles in their social contexts has not been studied before. To make the problem even more realistic for open societies, we do not assume that agents share knowledge on their common resources. So, they have to compute semantic agreements towards performing their joint actions. 
%The paper reports on an empirical study of whether  and how efficiently societies of agents converge to norms, exploring the proposed social learning processes w.r.t. different society sizes, and the ways agents are connected. The results reported are very encouraging, regarding the speed of the learning process as well as the convergence rate, even in quite complex settings.
\end{abstract}
\section{Introduction}
It is well known that effective norms, policies or conventions can significantly enhance the performance of  agents acting in groups and societies, since they do enable a kind of social control to the behavior of agents, without compromising their autonomy \cite{Eipstein}.  The emergence or learning of norms in agents' societies is  a major challenge, given that societies are open,  agents may not be qualified to collaborate effectively under previously unseen conditions, or they may need to compute effective rules of behavior very efficiently, w.r.t. their preferences and constraints.

There are several models that have been proposed for computing agreed and emerging norms / conventions via learning from agents' interaction experiences. The research question is when and how agents converge to  agreed norms or conventions, in cases there are multiple strategies yielding the same optimal payoff. 
In this article we consider norms as social conventions (i.e. as set of agreed, stipulated, or generally accepted standards or criteria) rather than as deontic aspects (e.g. obligations, prohibitions or permissions).The  main question that this article aims to answer is "how effectively do norms emerge in a society via establishing agreements in social contexts through local interactions and with limited information about others knowledge,  preferences and choices?".  

To frame the existing computational models towards the emergence of norms, as pointed out in \cite{NormLearning}, these may be categorized to imitation, normative advise, machine learning and data-mining models.  In this paper we propose reinforcement learning approaches to computing norms. Previous studies (e.g. \cite{MukherjeeSA08}, \cite{SenIJCAI}, \cite{Yu2013}), have shown that Q-learners are more efficient than other learners using for instance WoLF  \cite{Bowling02}, Fictitious Play \cite{Fudenberg}, Highest Cumulative Reward -based \cite{Shoham1997} models.  In this article we aim beyond the reported facts towards showing how agents mutually learn in a distributed and efficient way strategies that maximize their payoff w.r.t. some preferences and constraints, via local interactions, while playing multiple and maybe  roles with incompatible preferences. Towards this target, this article reports on social learning models where agents via communication do form specific expectations for the behavior of others. They do this in their social contexts, ensuring local interactions: Agents exchange their computed strategies, and  compute the best strategy to follow, given the strategies of others.  
Additionally, while some of the research works have considered local agents' interactions, and learning-by-doing via local interaction with multiple agents, this article emphasizes on the social context of agents. By doing so, we formalize the learning process in cases where agents play multiple roles simultaneously in their contexts and by interacting with acquaintances playing different roles. The notion of agents' social context allows distinguishing between agreements of agents in their social context and society-wide norms. These, in conjunction to relaxing the assumption that agents share a common representation of the world, clarify the contributions of this research work, compared to the state of the art approaches for learning norms. Specifically, this article proposes methods for agents to reach social conventions, by:
\begin{itemize}
\item Playing multiple roles and interacting with others that play multiple roles simultaneously in their social contexts.
\item Reconciling conflicting options while considering incompatibilities among roles.
\item Computing semantic correspondences between their representations of the world: This article deals with strategies on using resources towards performing joint tasks. Although the article considers a specific type of resource (time),  this is not restrictive to the applicability of the methods proposed to other type of resources are strategies.
\end{itemize}
Finally, 
\begin{itemize}
\item Agents learn society norms (conventions) via computing agreements in their social contexts. 
\end{itemize}
%, assuming that agents are connected with others in their social contexts, that in their own turn are connected to others, and so on and so forth. 

We need to emphasize that agents' semantic agreements (i.e. agreements on the meaning of terms they use for the representation of resources) are put in the context of their joint tasks: Tasks that need the coordinated action of at least two agents.  By doing so, agents are restricted to semantic agreements that do "work in reality" effectively. Consider for instance the case where agents, due to their limited knowledge of others' representations reach agreements about the meaning of symbols, which if put to a working context will lead either to ineffectiveness, or to the inability to act.

Towards answering the main research question stated above, this article proposes two social learning reinforcement learning methods. Both methods exploit (a) agents' preferences on the use of resources, (b) the feedback that agents receive for their strategies to use resources while interacting with others in their social context, (c) their reward for performing role-specific actions for any role they play w.r.t. their strategies for using resources. Although both methods are social (i.e. they are based on agents' local interactions), in one of them agents do not consider their joint decisions and do not share rewards, while in the other they learn collaboratively by acting and receiving rewards/sanctions for their joint strategies.

This article is structured as follows: Section 2  presents a motivating scenario. Section 3 formulates the problem and section 4 presents the proposed methods. Section 5 presents experimental results for agents societies of different size and structure. Section 6 presents related works and finally, section 7 concludes the article by summarizing the contributions made and presents interesting lines of future research.

\section{Motivating Scenario}
Consider the following scenario: AgentX among other commitments in its working context is being involved in a recently-appointed team towards the design of a new product. The team has a coordinator agent who has already many commitments with other teams. Besides that, AgentX  is committed to perform other tasks as member of other  groups independently from his working context. Role-specific tasks that AgentX has to perform require resources whose use has to be coordinated with others. Time is of primary importance here. Thus, to arrange his schedule, AgentX considers different time periods for scheduling role-specific tasks according to roles' preferences (e.g. due to conventional arrangements). He tries to keep concerns separate, while complying with his commitments and obligations and coordinating with others effectively. 

The social context of AgentX is the set of roles to which the agent interacts, including also own roles. The agents with whom he interacts may also play multiple roles, and they constitute AgentX's neighborhood. 

Consider for instance  two daily periods that AgentX names $P_1$ and $P_2$. Given a role $R$ that AgentX plays, these periods may be ordered according to a measure of preference: Let that be $\gamma(R,\cdot)$.  Let for instance $P_1$ be more preferred than $P_2$ from the point of view of role $R$. I.e. $\gamma(R,P_2) \leq \gamma(R,P_1)$. While AgentX's collaborators may play other roles (different from $R$) they may not share AgentX' representation of periods. Thus, AgentX has to agree with them on the meaning of periods, to start coordinating with them. Consider for instance the periods $X_1$ and $X_2$ specified by the busy team coordinator. Agents, to begin coordinating towards performing joint tasks (e.g. meeting), need to reach agreement to the correspondences between the periods considered by the different roles \footnote{Of course agents may specify periods using different time granularities, different forms of representing time, etc. In this paper we assume that there is a specific granularity for specifying periods and thus agents just have to align their specifications: Otherwise, further agreements are necessary.}. In this particular case there are clearly two possible options for reaching an agreement on correspondences between periods. Nevertheless, there is at most one option which is meaningful (i.e. it corresponds to the semantics of periods' representations), but we do not assume this to be known to the agents. Let, for instance, the meaningful correspondences be: (a) $P_1$ is the same as $X_2$ and  (b) $P_2$ is the same as  $X_1$. Notice that agents may reach an agreement to non-meaningful correspondences: In this case they will not be able to act jointly, receiving a very low payoff for their joint task.

Notice that to reach an agreement to the correspondences between periods' representations agents do not have to consider their preferences on periods for scheduling tasks.
%This can be modeled as a  social agreement game using the payoffs in Table \cite{socialagreement} 
%\begin{center}
% \begin{tabular}{  c  |  c | c | }
%    & $P^{member}_1$ & $P^{member}_2 \\ \hline
%    $X^{coordinator}_1$ & -1,-1 & 3,3 \\ \hline
%   $X^{coordinator}_2$ & 3,3 & -1,-1 \\
%    \hline
%    \label{socialagreement}
%  \end{tabular}
%\end{center}
Nevertheless, considering preferences, AgentX's neighbors may have the incentive of choosing AgentX's non-preferable period for scheduling their tasks. Such decisions can lead to undesirable situations and to ineffectiveness in performing joint tasks. Consider for instance the busy team coordinator. He/she may prefer  $X_1$ to schedule meetings with team members, while AgentX -being a member of the team, prefers to schedule joint tasks in period $P_1$. Given, for instance, that $P_1$ is the same as $X_2$ and  $P_2$ is the same as  $X_1$, then, the possible choice of the team member to schedule a joint task is  $P_1$ (and $X_2$), while for the team coordinator is $X_1$ (and $P_2$): These possible choices do not satisfy the preferences of both agents. 
%It must be noticed that, given an agreement on correspondences that are not meaningful (e.g. $P_1$ with $X_1$ and  $P_2$ with  $X_2$), agents will not be able to perform their joint tasks since the corresponding periods are actually disjoint.

%This is modeled as a social dilemma using the payoffs in Table \cite{socialdilemma1}.  The preferences can be omitted in Table \cite{socialdilemma1} since they are specified by the payoffs.

%\begin{center}
%  \begin{tabular}{  c  |  c | c | }
%    & $P^{member}_1$ & $P^{\widetilde{member}}_2 \\ \hline
%    $X^{\widetilde{coordinator}}_1$ & -1,-1 & 2,3 \\ \hline
%    $X^{coordinator}_2$ & 3,2 & -1,-1 \\
%   \hline
%    \label{socialdilemma1}
%  \end{tabular}
%\end{center}

 %\begin{center}
 % \begin{tabular}{  c  |  c | c | }
 %    & $P^{member}_1$ & $P^{\widetilde{member}}_2 \\ \hline
 %   $P^{\widetilde{family_member}}_1$ & -1,-1 & 2,2 \\ \hline
 %   $P^{family_member}_2$ &  2,2 & -1,-1 \\
 %   \hline
  %  \label{socialdilemma2}
 % \end{tabular}
%\end{center}

In addition to these, some of the roles may have incompatible requirements and preferences to the use of  resources. We define two roles to be \textit{incompatible w.r.t. a resource} (or simple $incompatible$, in case we consider time as the only type of resources) if joint tasks for these roles cannot  share the resource when performed by a single agent (e.g. considering time periods, an agent must schedule tasks for two incompatible roles in non-overlapping time periods). 

Thus, summarizing  the above, AgentX has to reach agreements with his neighbors to schedule their joint tasks, so as to satisfy as much as possible his preferences on scheduling tasks, and the constraints related to the incompatibility of roles: This is rather complicated given that AgentX plays multiple roles and interacts with multiple others, while this is true for his acquaintances.

For a convention to evolve in the society, all agents in the population playing the same roles have to agree on their strategies for using resources: E.g. pairs of agents playing the roles of team members and team coordinators have to learn one of the following policy pairs to schedule joint tasks: (a) ($P_1$, $X_2$),  according to the preference of the team member, or (b) ($P_2$, $X_1$), being in accordance to the preferences of the coordinator.

This scenario emphasizes on the following aspects of the problem:
\begin{itemize}
\item Related to resources:
	\begin{itemize}
	\item Agents need to coordinate their use of resources to perform joint tasks  (in our scenario we consider time as the unique resource)
	\item Agents do not share a common representation of the resources, so they have to agree on the semantics of their representations
	\item Agents'  preferences on the use of the resource vary for each of the roles they are playing
	\end{itemize}	
\item Related to agents' roles:
	\begin{itemize}
	\item Each agent may play and interact with multiple (even incompatible) roles.
	\item Each agent has a social context, defined by its own roles and the roles that it interacts with.
	\end{itemize}
\item Related to agreements and norms:
	\begin{itemize}
	\item Semantic agreements are put in the context of agents actions: In our example it is clear that even if agents agree on correspondences between periods, this may not lead them to schedule their tasks as effectively as they may wish.
	\item Agents in their social context have to reach agreements on the use of resources for performing their joint tasks.
	\item Norms are agreements that are widely accepted by all agents in the society.
	\end{itemize}
\end{itemize}

As far as we know, there is not any research work concerning the emergence of conventions in agents' societies that consider these  aspects in combination. As already said, the major question that this paper aims to answer is "how effectively do norms emerge in a society via establishing agreements in social contexts through local interactions and with limited information about others' representations, preferences and choices?". The effectiveness of a model is measured by means of the percentage of role playing agents reaching agreement on specific conventions, as well as by measuring the computational iterations (epochs) necessary for a society to converge to conventions.

Towards answering this question, agents need to (a) compute semantic agreements for the terms they use to represent resources, (b) use semantic agreements to  compute agreements on the use of resources for performing their joint tasks in their social contexts w.r.t. their preferences on using resources and roles' incompatibilities. 
 
\section{Problem Specification}
A society of agents $S=(\mathcal{R}, A,E)$ is modeled as a graph with one vertex per agent in $A$ and any edge in $E$ connecting pairs of agents. A connected pair of agents must be coordinated to the use of resources for the performance of role-specific tasks (e.g. to the scheduling of their tasks) and can communicate directly to each other.  Each agent $i$ in the society is attributed with different roles $\mathcal{R}=\{R_1,  R_2...\}$. The naming of roles is a social convention and thus, all agents in the society use the same set of roles. $N(i)$ denotes the neighborhood of agent $i$, i.e. the set of agents connected to agent $i$, including also itself. Subsequently, the fact that agent $i$ plays the role $R_j \in \mathcal{R}$, is denoted by \textit{i:j}.

Each role $R_i$ considers a set of  time periods $\mathcal{P}_{R_i}=\{P_1,  P_2...\}$ that are ordered according to $R_i$'s preferences for scheduling role-specific tasks.  Role-specific periods  in $\mathcal{P}_{R_i}$ are order by the preference of  $R_i$, according to the function $\gamma(R_i,\cdot): \mathcal{P}_{R_i} \rightarrow \mathbb{R}$. Although we may consider any relation between periods (e.g. they may be disjoint, overlapping etc), in this article we consider only equal (=) and mutually disjoint ($<>$, non-overlapping) time periods.  Each role has its own preferences to scheduling tasks in periods, while the naming of periods as well as the pairs of incompatible roles  is common knowledge to all agents that play the same role.

Given a pair of roles ($R_i$, $R_j$), these may be $incompatible$ w.r.t. a resource. Considering time, agents interacting with incompatible roles cannot schedule any pair of joint tasks, with each these roles, during the same time period. Any pair of agents, or a single agent, may play incompatible roles.

Agents playing different roles do not possess any common knowledge, neither exchange any information concerning the role-specific periods, their preferences on scheduling tasks, or their payoffs for scheduling tasks in any period. Thus, agents  playing different roles may use different names for the same period, or the same name for denoting different periods. No agent possesses global knowledge on the semantics of role-specific representation of periods , and thus on  correspondences between periods names: We consider that this holds for any single agent that plays multiple roles, as well. 

At this point it must be emphasized that while this article considers time periods, the formulation and the proposed methods can be applied to other types of resources that can be treated similarly to time and are necessary to the execution of role-specific tasks.  

A social context  for an agent $i$ denoted by $SocialContext(i)$, is the set of roles played by the agents in its neighborhood. More formally: 

$SocialContext(i)= \{R_k | \exists j \in N(i) \textit{ and } j:k \}.$ 

It must be noticed that the social context of an agent $i$ includes own roles, denoted by $Roles(i)$. %It must be noticed that if $R_k \in SocialContext(i)$, then $R_m \in SocialContext(j)$, for any $j \in N(i)$ and for roles $R_k, R_m$, s.t. $i:m$ and $j:k$.

Agents in the society must decide on the scheduling of their (more interestingly, joint) tasks so as to increase their effectiveness. More specifically, considering two acquaintances  \textit{i:k} and  \textit{j:m}, where $j \in N(i)$, and a joint task for their roles $R_k$ and  $R_m$,  agents must schedule that task in an agreed period $P$, so as to increase their expected payoff with respect to their role-specific preferences on schedules. Considering that agents and their neighbors play multiple -maybe incompatible- roles,  they have also to take into account  role-specific (incompatible) requirements on scheduling tasks. Incompatibilities are formally specified in section 3.

To agree on a specific period $P$ for scheduling their joint task, agents $\textit{i:k}$ and $\textit{j:m}$ have to first agree on correspondences between their representations of periods: Towards this we consider that agents can subjectively hold correspondences between own representations of periods and representations of others: These may be computed by each agent using own methods, and information about others' roles. 
A subjective correspondence for the agent \textit{i:k} and its acquaintance \textit{j:m} is a tuple $\langle P, S \rangle$, s.t. $P \in \mathcal{P}_{R_k}$ and $S \in \mathcal{P}_{R_m}$\footnote{It must be pointed out that  since the neighborhood of any agent includes itself, and its social context includes its own roles, it may also hold that $i=j$.}. Such a correspondence represents that the agent $i$ considers $P$ and $S$ to represent the same time interval. Nevertheless, given that acquaintances may nor agree on their subjective correspondences, they have to reach an agreed set of correspondences.

For norms to emerge in the society, any pair of agents (anywhere in the society) playing roles $R_k, R_m$ must reach the same decisions for scheduling joint tasks for these roles. 
 
Towards this goal, this article proposes two distributed social learning methods for agents to compute society-wide agreements via local interactions with their neighbors. 
\section{Social Reinforcement Learning  methods for computing Agreements}
To describe the proposed methods for the computation of norms, we distinguish between two, actually highly intertwined, computation phases: (a) The computation of agent-specific, subjective correspondences on periods, and strategies for scheduling tasks w.r.t. own preferences and constraints concerning incompatibility of roles; and (b) the computation of contextual agreements concerning agents' strategies to schedule joint tasks. 
%Finally, we show how do norms emerge via the computation of agreements in agents' social contexts.
\\ \\
\textbf{Computation of local correspondences and strategies:}
Given an agent $i$ playing a role $R_k$,  and a role $R_m \in SocialContext(i)$ played by a an agent $j$ in the neighborhood of $i$, agents need to compute subjective correspondences between periods in $\mathcal{P}_{R_k}$ and $\mathcal{P}_{R_m}$.

%, and  $\mathcal{P}_{R_m}$ be the set of time periods known by any $R_m$-playing agent.  Of course, agents prefer to schedule tasks in the most preferred periods according to the preferences $\gamma(R,P)$, where $P \in \mathcal{P}_{R}, R \in \mathcal{R}$. 

Although agents may use own methods to compute these correspondences, these computations have to preserve the semantics of periods' specifications: This is done via validity constraints that coherent  correspondences between periods must satisfy. These constraints depend on the possible relations between periods.  Therefore, considering only equal and disjoint time periods, and given  two distinct roles $R_k$ and $R_m$, the validity constraints that correspondences  computed by \textit{i:k} must satisfy are as follows:
\small
\begin{itemize}
\item if $\langle P, X\rangle$ and $\langle P', X'\rangle$ are correspondences with $X,X' \in  \mathcal{P}_{R_m}$, $P,P' \in  \mathcal{P}_{R_k}$ and $P <> P' $, then it must hold that $X <> X'$.
\item if $\langle P, X\rangle$ and $\langle P, X'\rangle$ are correspondences with $X,X' \in  \mathcal{P}_{R_m}$ and $P,P' \in  \mathcal{P}_{R_k}$, then $X = X'$.
\end{itemize}
\normalsize
Given these validity constraints, each agent can compute its own role-specific, subjective, coherent correspondences between time periods. Given these correspondences, any agent $\textit{i:k}$ has to make a specific decision for the period  to schedule joint tasks with any other agent playing the role $R_m$ in its social context.  Let that decision be denoted by $decision(\textit{i:k},\textit{$\cdot$:m})$ \footnote{the notation ($\textit{$\cdot$:m}$) means "any agent playing the role $R_m$"}.  Later on we specify how agents reach these decisions and how they reach agreements on their subjective correspondences. 

Given that each agent may interact with multiple roles in its social context, considering any  pair of incompatible roles $R_k,R_m$, the following incompatibility constraint holds:
\small
\begin{itemize}
\item Given an agent $i$ playing any role $R_x$,  and given two incompatible roles $R_k,R_m \in SocialContext(i)$, then \\ $decision(i:x,\textit{$\cdot$:m}) <> decision(i:x,\textit{$\cdot$:k})$. 
\end{itemize}
\normalsize
%It must be pointed out that validity constraints hold for any pair of roles played by any pair of agents in a social context, while incompatibility constraints hold for any pair of incompatible roles played by a single agent.
Given the above validity and incompatibility constraints, the utility of an agent $i:k$ for choosing a  period $P \in P_{R_k}$ to schedule joint tasks with $j:m$, given the subjective correspondence $<P,X>$ between periods, is  \\
\small
$U(\textit{i:k},P)=\gamma(R_k,P)+f(\textit{i:k},P)$, where 
$\gamma(R_k,P)$ is the preference of role $R_k$ to $P$, and 
$f(\textit{i:k},P)= G(\textit{i:k})+C(\textit{i:k})$, where 
$G(\textit{i:k})=Payoff*SatisfiedConstraints(\textit{i:k})$ and \\ $C(\textit{i:k})=Penalty*ViolatedConstraints(\textit{i:k})$.
% \left\{
%\begin{tabular}{l l}{\columnwidth}%{lp{5cm}}
%$G$ & if all constraints related to $R_k$ \\
%& in the neighborhood of $i$ are satisfied \\ 
%& $G=Payoff*SatisfiedConstraints$ \\
%$C$ & in any other case, \\ 
%& where $C=Penalty*ViolatedConstraints$.
%\end{tabular} \right\}
%\end{math}
%\end{itemize}
\normalsize

$Payoff$ is a positive number representing the payoff of any satisfied constraint in the social context of agent \textit{i:k} and $Penalty$ is a  negative number that represents the cost of violating a validity or incompatibility constraint. $SatisfiedConstraints(\textit{i:k})$ (resp. $ViolatedConstraints(\textit{i:k})$) is the number of satisfied (resp. violated) constraints for the agent $i$. 
\\ \\
\textbf{Computing contextual agreements:}
Given agents' subjective correspondences and own decisions for any role they play, these correspondences and decisions may not agree with the choices of their neighbors. Towards reaching agreements, also with respect to constraints and role-specific preferences, agents consider the feedback received from their neighbors.

According to this communication-based learning approach, given an agent $i$ and two roles $R_k \in Roles(i)$ and $R_m \in SocialContext(i)$,  to get feedback on decisions, the agent $\textit{i:k}$ propagates its decision for scheduling joint tasks with  agents $\cdot\textit{:m}$ in its neighborhood in period $P$, together with its subjective correspondence  $\langle P, X \rangle$, where $X \in \mathcal{P}_{R_m}$ to all $R_m$-playing agents in $N(i)$. 
It must be noticed that the propagated decision concerns a specific pair of role playing agents and both, a period and a subjective correspondence for this period. Such a decision is of the form $(\textit{i:k}, \textit{x:m}, \langle P, X\rangle)$, where $x \in N(i)$ and $decision(\textit{i:k}, \cdot \textit{:m})=P$.

Agents propagate their decisions to their neighbors in the network iteratively and in a cooperative manner, aiming to exploit the transitive closure of correspondences in cyclic paths. This is similar to the technique reported in \cite{Vouros2013}. Agents propagate what we call $c-histories$, which are ordered lists of decisions made by agents along the paths in the network. Each propagated decision heads such a history. For instance the c-history propagated by $i$ to any $R_m$-playing agent $x$, as far as the role $R_k$ is concerned, is [$(\textit{i:k}, x:m, \langle P, X \rangle) |L$], where $L$ is either an empty c-history or the c-history that has been propagated to $i$, concerning its role $R_k$. 
By propagating c-histories, agents can detect cycles and take advantage of the transitivity of correspondences, detecting positive/negative feedback to their decisions. 

Specifically, an agent $i$ detects a cycle by inspecting in a received c-history the most recent item $(\textit{i:k}, x:m, \langle P, X\rangle)$ originated by itself: Given a cycle $(1 \rightarrow 2 \rightarrow... (n-1) \rightarrow 1)$, then for each  decision  $(\textit{1:k}, \textit{2:m}, \langle P, X \rangle)$ for the roles $R_k$ and $R_m$ that agents $1,2$ play, respectively, heading a c-history from  $1$ to $2$, the originator must get a decision $(\textit{n-1:m}, \textit{1:k},\langle P, X \rangle)$ from the last agent $(n-1)$ in the cycle, if it plays the role $R_m$. Thus, the agent $1$ must receive a decision from $(n-1)$ concerning $P$, rather than to any other period, and the correspondence $\langle P, X \rangle$. In such a case the agent $1$ counts a positive feedback. In case there is a cycle but the forwarded decision does not concern $P$, then there are one or more correspondences or decisions through the path  that result to disagreements. In this case, the agent $1$ counts a negative feedback for its decision.  It must be noticed that disagreements may still exist when the agent  $1$ gets the expected choice but several decisions along the path compensate "errors". These cases are detected by the other agents, as the c-history propagates in the network.  
To make the computations more efficient and in order to synchronize agents' decision making we consider that c-histories can be propagated up to 3 hops with repetitions: This means that given two neighbors $i$ and $j$, any c-history starting from $i$ (1st hop) shall be returned to this agent with the decision of $j$ (2nd hop), and will return later to $j$  with the new decision of $i$ (3rd hop). In the last hop the agent $i$ will choose a strategy by considering also the feedback received from $j$, in conjunction with feedback from any other neighbor. 

But how actually do agents compute decisions in their social context w.r.t. their preferences and constraints? Notice that  decisions concern specific periods w.r.t. subjective correspondences. From now on, when we say \textit{decisions} we  mean exactly this combination: Thus when agents revise their decisions they may revise their subjective correspondences, or their strategies for scheduling tasks, or both. 
\\ \\
\textbf{Reinforcement learning and the emergence of norms:}
Given that agents do not have prior knowledge about the effects of decisions made, this information has to be learned based on the rewards received (including feedback from others).  Using the model of collaborative multiagent MDP framework \cite{PutermanMDP}, \cite{GuestrinPhD} we assume: \\
-The society of agents $S=(\mathcal{R},A,E)$.\\
-A time step $t=0,1,2,3,...$ \\
-A set of discrete state variables per agent-role $\textit{i:k}$ at time $t$, denoted by $s^t_{(\textit{i:k}),(\textit{$\cdot$:m})}$, where $i \in A$ and $R_m \in SocialContext(i)$. The state variable ranges to the set of possible correspondences between periods in $P_{R_k}$ and periods in $P_{R_m}$. The local state $s^t_{i}$ of agent $i$ at time $t$ is the tuple of the state variables for all roles played by $i$ in combination with any role in  its social context. A global state $s^t$ at time $t$ is the tuple of all agents' local states. The set $State$ is the set of global states. \\
 %\footnote{The notation is as follows: For  states/decisions/rewards of agent $\textit{i:k}$ considering   interactions with $R_m$-playing agents, we use subscripts of the form ($\textit{i:k},\textit{j:m}$), for agents local states/decisions/rewards for all their roles we use ($i$) and for joined states/decisions/rewards in the society there is no subscript.}
-A strategy for every agent-role $\textit{i:k}$ and role $R_m \in SocialContext(i)$ at time $t$, denoted by $c^t_{(\textit{i:k}),(\textit{$\cdot$:m})}=decision(\textit{i:k},\textit{$\cdot$:m})$. The local strategy for every agent $i$, denoted by $c^t_i$ is a tuple of strategies, each for any role that $i$ plays in combination with any other role in its social context. The joint strategy of a subset $T$ of $A \times \mathcal{R}$ (for instance of agents in $N(i)$ playing their roles in $SocialContext(i)$), is a tuple of local strategies, one for each agent playing a role in that set, denoted by $c^t_T$ (e.g. $c^t_{N(i)}$). The joint strategy for all agents $A$ at time $t$ is denoted $c^t$, while the set of all joint strategies for $A$ is the set $Strategy$. \\
-A state transition function $T: State \times Strategy \times State \rightarrow [0,1]$ gives the transition probability $p(s^{t+1}|s^t,c^t)$, based on the joint strategy $c^t$ taken in state $s^t$. \\
-A reward function per agent-role $\textit{i:k}$ given its decisions concerning role $R_m \in SocialContext(i)$, denoted by $Rwd_{(\textit{i:k}),(\textit{$\cdot$:m})}$, where $i \in A$ and $R_k$ a role played by agent $i$. The  reward function per agent-role $\textit{i:k}$, denoted by $Rwd_{(\textit{i:k})}$ provides  the agent $\textit{i:k}$ with an individual reward based on the joint decision of its neighborhood, taken in its local state. The local reward of an agent $i$, $Rwd_{i}$, is the sum of its rewards for all the roles it plays.    

It must be noticed that states represent agents' assumptions about periods' correspondences, while agents' strategies concern the specific periods for scheduling role-specific tasks. The reward function concerns decisions made by agents, i.e. agents' strategies w.r.t. their states, and depends on the utility of agents' choices while playing specific roles, on the feedback received from neighbors, and on the payoff received after performing the scheduled tasks:
\\
\small
\noindent $Rwd_{(\textit{i:k})}(P,s_i) = a*U(\textit{i:k},P)+b*Feedback(\textit{i:k},s_i)+Payoff(\textit{i:k},c_i)$, \normalsize where \small 
\\
\noindent $Feedback(\textit{i:k},s_i)=Payoff*Feedback^+ (\textit{i:k},s_i)+Penalty*Feedback^-(\textit{i:k},s_i)$,
\normalsize
\\ $P \in \mathcal{P}_{R_i}$, and $Feedback^+(\textit{i:k},s_i)$, $Feedback^-(\textit{i:k},s_i)$ are the numbers of positive and negative feedbacks received, respectively, $Payoff$ and $Penalty$ are the numbers specifying the payoff and cost for each positive and negative feedback, respectively (being equal to the corresponding utility parameters).
The parameters $a$ and $b$ have been used for balancing between own utility and feedback received by others: As previous works have shown  \cite{Vouros2013}, the role of both is crucial. The method is tolerant to different values of these parameters, but here we consider that $\frac{a}{b}=\frac{1}{10}$.  Finally, $Payoff(\textit{i:k},c_i)$ is the payoff that the agent $i$ receives after performing $R_k$ tasks by applying the strategies chosen. 

A (local) policy of an agent $i$  in its social context is a function $\pi_i: s_i \rightarrow c_i$ that returns a local decision for any given local state. The objective for any agent in the society is to find an optimal policy $\pi^*$ that maximizes the expected discounted future return $V_i^*(s) = max_{\pi_i}E[\sum\limits_{t=0}^{\infty} \delta^tRwd_i(\pi_i(s^t_i), s^t_i)|\pi_i)]$ for each state $s_i$, while playing all its roles. The expectation $E(.)$ averages over stochastic transitions, and $\delta \in [0,1]$ is the discount factor. 

This model assumes the Markov property, assuming also that rewards and transition probabilities are independent of time. Thus, the state next to state $s$ is denoted by $s'$ and it is independent of time.

$Q-$functions, or action-value functions, represent the future discounted reward for a state s when making the choice c and behaving optimally from then on. The optimal policy for the agents in state s is to jointly make the choice $argmax_cQ^*(s,c)$ that maximizes the expected future discounted reward.  %Reinforcement learning can be applied to estimate $Q^*(s,a)$, when agents do not know the transition and reward model. 
The next paragraphs describe two distributed variants of  Q-learning considering that agents do not know the transition and reward model (model-free methods) and interact with their neighbors, only.
Both variants assume that agents propagate their decisions to neighbors, and % Each local function $Q_i$ concerns only agents' local states and choices (i.e. states and choices concerning agents social context).
% $Q(s,c) = \sum\limits_{i=1}^{|A|} Q_i(s_i, c_i)$,
% where $s_i$ and $c_i$ are respectively the local state and local choice for agent $i$. 
take advantage of  dependencies with others, specified by means of the edges connecting them in the society.  
\\ \\
\textit{Independent Reinforcement Learners:} In the first variant, the local function $Q_i$  for an agent $i$ is defined as a linear combination of all contributions from its social context, for any role $R_k$ played by $i$ in combination with roles  $R_m$ in its social context: $Q_{i}=\sum\limits_{R_k} \sum\limits_{\textit{j:m}, j\in N(i)} Q_{(\textit{i:k}),(\textit{j:m})}$. To simplify the formulae we denote ($(\textit{i:k}),(\textit{j:m})$) by $i\bowtie j$. Thus, each $Q_{i\bowtie j}$ is updated as follows:

\small
$Q_{i\bowtie j}(s_i,c_{i\bowtie j})= Q_{i\bowtie j}(s_i,c_{i\bowtie j})+ \\ \alpha [Rwd_{\textit{i:k}}(s_i,c_{i\bowtie j})+\delta max_{c'_{i\bowtie j}}Q_{i\bowtie j}(s_i',c_{i\bowtie j})-Q_{i\bowtie j}(s_i,c_{i\bowtie j}))]$
\normalsize  

This method is in contrast to the Coordinated Reinforcement Learning model proposed by Guestrin in \cite{GuestrinCRL} that considers society's  global state, and it is closer to the model of independent learners, since the formula considers the local states of agents.  
  \\ \\
\textit{Collaborative Reinforcement Learners:} The second variant is the agent-based update sparse cooperative edge-based Q-learning method proposed in \cite{Kok2006}. Given  two neighbor  agents $\textit{i:k}$ and $\textit{j:m}$, the $Q-$function is denoted $Q_{\textit{i:k},\textit{j:m}}(s_{\textit{i:k},\textit{j:m}},c_{\textit{i:k},\textit{j:m}},c_{\textit{j:m},\textit{i:k}})$, or succinctly $Q_{i\bowtie j}(s_{i\bowtie j},c_{i\bowtie j},c_{j\bowtie i})$, where $s_{i\bowtie j}$ are the state variables related to the two agents playing their roles, and $c_{i\bowtie j},c_{j\bowtie i}$ are the strategies chosen by the two agents. The sum of all these edge-specific $Q-$functions defines the global $Q-$function. It must be noticed that it may hold that $i=j$, considering the $Q-$functions for the different roles the agent $i$ is playing.
The update function is as follows:
\\
$Q_{i\bowtie j}(s_{i\bowtie j},c_{i\bowtie j},c_{j\bowtie i}))= Q_{i\bowtie j}(s_{i\bowtie j},c_{i\bowtie j},c_{j\bowtie i})) + \\ \alpha \sum\limits_{x:y \in \{\textit{i:k}, \textit{j:m}\}}  \frac{Rwd_{x:y}(s_{x:y},c_{x:y}) + \delta Q^*_{x:y}(s_{x:y}',c_{x:y}) -  Q_{x:y}(s_{x:y},c_{x:y}) }{|N(x)|} $
\\
The local function of an agent $\textit{i:k}$ is defined to be the summation of half the value of all local functions $Q_{i\bowtie j}(s_{i\bowtie j},c_{i\bowtie j},c_{j\bowtie i})$ for any \textit{j:m}, with $j \in N(i)$ and $R_m \in SocialContext(i)$: \footnotesize
$Q_{\textit{i:k}}(s_{\textit{i:k}},c_{\textit{i:k}}) = \frac{1}{2} \sum\limits_{j:m} Q_{i\bowtie j}(s_{i\bowtie j},c_{i\bowtie j},c_{j\bowtie i})$.
\normalsize
\begin{figure}
\centering
\includegraphics[width=.40\textwidth]{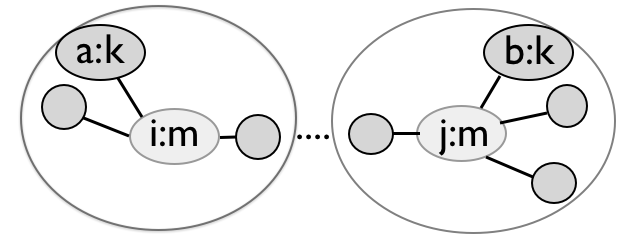} 
\vspace{-9pt}
\caption{Isolated agents playing the role $R_k$.}
\vspace{-10pt}
%\ref{exps}
\end{figure}
Closing this section we need to answer whether agents in any society do learn social norms via agreements in their social context: The answer is negative in case there are socially-isolated agents playing the same role. These are agents whose social context is limited to a single role. Thus, they do not interact "heavily" with the society and are somehow isolated in the neighborhoods of others.  These are for instance the agents $a$ and $b$ in Figure 1: They interact only with the agents \textit{i:m} and \textit{j:m}. Although \textit{i:m} and \textit{j:m} may reach agreements on their role-specific strategies via the path(s) connecting them, and each one of them may reach agreements with \textit{a:k} and \textit{b:k} in their social contexts, respectively, there may not be an agreement between $a$ and $b$, and thus a norm may not emerge for $R_m$-playing agents. Nevertheless, these agents do have separate concerns and have reached agreements in their contexts. Such cases do not exist in the experimental cases considered in the section that follows.
\section{Experimental Results}
We have performed simulations using the two social learning methods proposed in two types of networks: Small-world networks that have been constructed using the Watts-Strogatz model (W) \cite{WattsStrogatz}, and scale-free networks constructed using the Albert -Barab\'{a}si (B) model \cite{Barabasi}. For both types of networks we have experimented with different populations of agents, and with various degrees of agents' connectivity. For these types of networks, we have run experiments with populations of 10,20,50,100 and 200 agents, and with and average number of neighbors (ANN) 4,10,16,20. Each case is denoted by $X\_|N|\_ANN$, (e.g. $B\_100\_10$) where $X$ the network construction model. This article reports on results with B networks  with different $|N|$ and ANN=4, on results with B networks  with $|N|=100$ and different ANNs, and finally on W$^{0.5}$ networks with $|N|=100$ and different ANNs.

The society roles $\mathcal{R}$ are 4, $\mathcal{R}=\{fmember,worker,dependent,boss\}$ and each agent can play up to 2 roles satisfying the following constraints: Any $worker$ can be an $fmember$ and vise-versa, a $dependent$ can not play any other role, while a $boss$ cannot play other roles and is connected to agents playing the role of $worker$.  The $dependents$ are up to 10\% of the population. Using these constraints, roles are assigned to agents randomly. The incompatible pairs of roles are ($fmember, worker$), ($fmember, boss$), ($dependent, worker$), ($dependent, boss$). Thus, any agent connected to an $fmember$ and a $boss$, for instance, can not schedule tasks for these two roles during the same period.

We do provide results when all agents are Independent Reinforcement-Learners (IRL) or Collaborative Reinforcement-Learners (CRL). In both methods the payoff $Payoff$ for positive feedback and satisfaction of constraints is equal to 3, while the penalty $Penalty$ is equal to -5.  Considering the reward, as already said, the ratio between the utility factor $a$ and the feedback factor $b$ is 1:10. For each role there are two distinct periods: The preferred ($p$) and the non preferred ($np$). These are denoted by the initial role of the role and a subscript $p$ or $np$. For instance $w_p$ is the workers preferred period. The joint task that agents need to perform is scheduling their meetings. The payoff matrices for role-specific strategies are given below.

\begin{center}
\small
\begin{tabular}{ c || c }
\begin{tabular}{ c | c | c }
&
$b_p$
&   
$b_{np}$ \\ \hline 
$w_p$ 
&   -1,-1
&  3,2 \\ 
$w_{np}$ 
& 2,3
& -1,-1 \\ 
\end{tabular}  
&
\begin{tabular}{ c | c | c }
&
$m_p$
&   
$m_{np}$ \\ \hline 
$w_p$
&   2,3
&  -1,-1 \\ 
$w_{np}$
& -1,-1
& 3,2 \\ 
\end{tabular}  \\ \\

\begin{tabular}{ c | c | c }
&
$d_p$
&   
$d_{np}$ \\ \hline 
$w_p$
&   3,3
&  -1,-1 \\ 
$w_{np}$
& -1,-1
& 3,3 \\ 
\end{tabular}  

&\begin{tabular}{ c | c | c }
&
$d_p$
&   
$d_{np}$ \\ \hline 
$m_p$
&   3,3
&  -1,-1 \\ 
$m_{np}$
& -1,-1
& 3,3 \\ 
\end{tabular}
\end{tabular}
\normalsize
\end{center}

It must be noticed that agents play different types of games while interacting with other roles, and do not exploit the payoffs of others in their neighborhood.
Both learning methods use an exploration function, counting the number of times each correspondence or strategy has been used. An epoch comprises an exploration followed by a pure exploitation period, while the number of times that correspondences and strategies are to be tried increases by a constant in each epoch.  

Figure 2 shows the results of both methods for different types of networks and different percentages of converging agents (T): The first (second) column reports on methods convergence when $T=100\%$ (respectively, when $T=90\%$). It must be noticed that results concerning state of the art methods require that $T \leq 90\%$.  The convergence rule is that the required percentage of agents has reached agreement without violating any constraint in 10 subsequent rounds during an exploitation period. 
Each point in any line is the average total payoff in 5 independent runs per case received by the agents at the end of an epoch. The reported results concern 9 epochs (1000 rounds), aiming to show the efficacy of the proposed methods. A line in Figure 2 stops at an epoch (notice that in some cases the X-axis has less than 9 points), when the corresponding method has converged in all independent runs for the corresponding case until this epoch. The average convergence round per case and method are reported in Figure 3. The value 1000 means that the corresponding method has not managed to  converge until epoch 9 (the 1000th round).

Experimentation results show that both methods are very effective both in agents convergence rate (i.e. percentage of agents reaching agreement) and in the number of epochs required. All cases converge, and in case we require  90\% convergence, agents using any of the methods managed to  converge to agreements in fewer than 9 epochs, except in networks with low ANN. Specifically, regarding the B networks with different populations (first two rows), as it is expected, the convergence is slower as the population increases. For networks of 100 agents, with a varying ANN, IRL converges faster for networks with higher ANN, while CRL is not affected by the degree of agents connectivity, although it converges slower than IRL in most cases when $T=90\%$. For W networks, both methods converge less effectively. However, CRL manages to convergence  more effectively when 90\% convergence is required, although this is not always the case: We can observe that in networks with a large population of agents and with high ANN, IRL can be more efficient. This is reported in all cases (especially for $T=90\%$) for B networks, but not for W networks.
\begin{figure}
  \vspace{-27pt}
\begin{tabular}{ l l }
\small
100\% convergence
&   
\small
90\% convergence \\ \hline \\ 
\includegraphics[width=.40\textwidth]{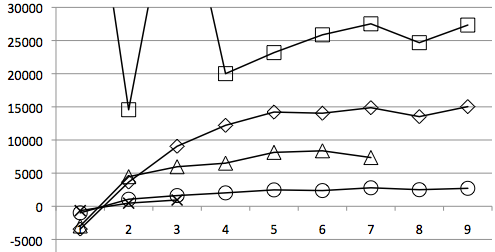} 
 &   
\includegraphics[width=.40\textwidth]{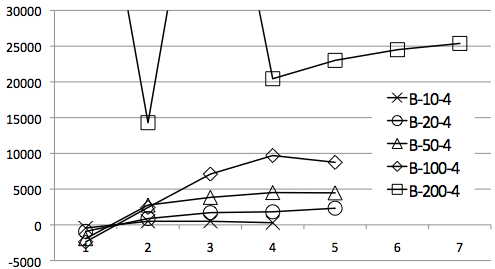}
\\ \small CRL, B, $|N| \in \{10,20,50,100,200\}$\\ 
\includegraphics[width=.40\textwidth]{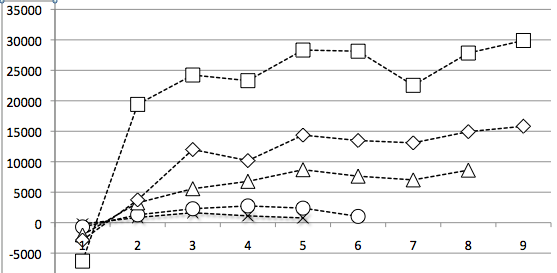}
 & 
\includegraphics[width=.40\textwidth]{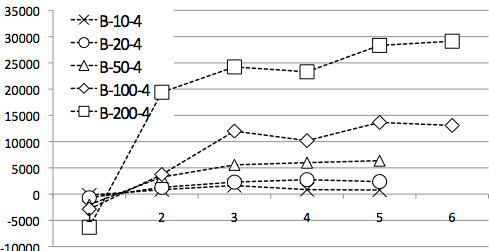}
\\ \small IRL, B, $|N| \in \{10,20,50,100,200\}$\\ 
\includegraphics[width=.40\textwidth]{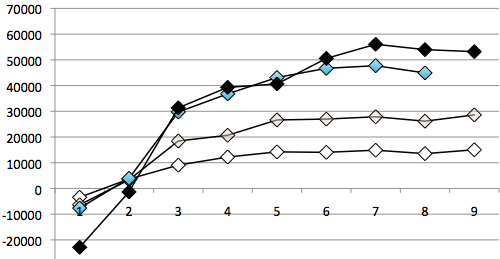}
 & 
\includegraphics[width=.40\textwidth]{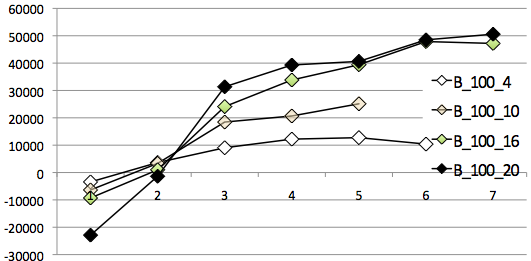}
\\ \small CRL,B,$|N|=100,\small ACD \in \{4,10,16,20\}$\\ 
\includegraphics[width=.40\textwidth]{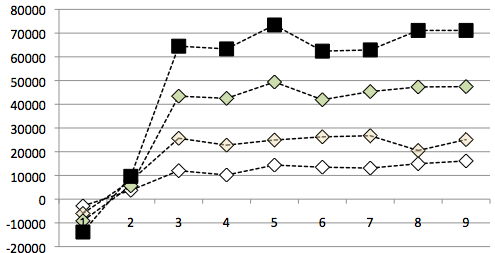}
 & 
\includegraphics[width=.40\textwidth]{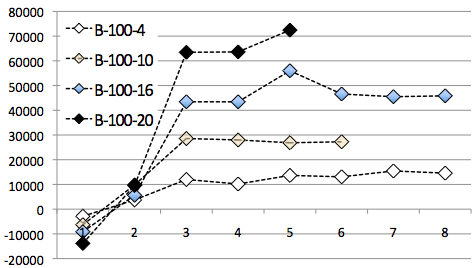}
\\ \small IRL,B,$|N|=100,\small ACD \in \{4,10,16,20\}$\\ 
\includegraphics[width=.40\textwidth]{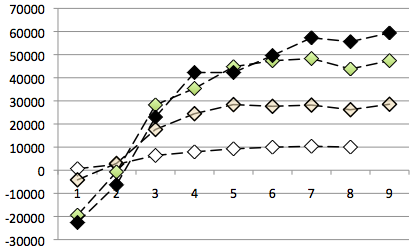}
 & 
\includegraphics[width=.40\textwidth]{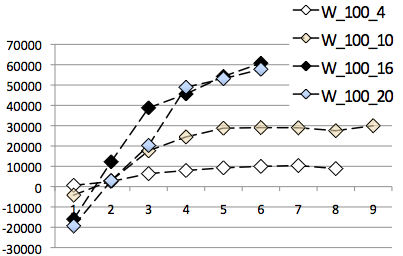}
\\ \small CRL,W,$|N|=100,\small ACD \in \{4,10,16,20\}$\\ 
\includegraphics[width=.40\textwidth]{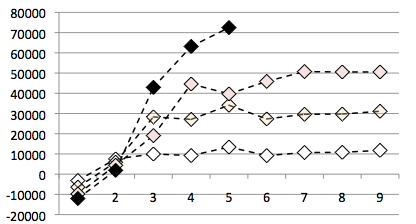}
 & 
\includegraphics[width=.40\textwidth]{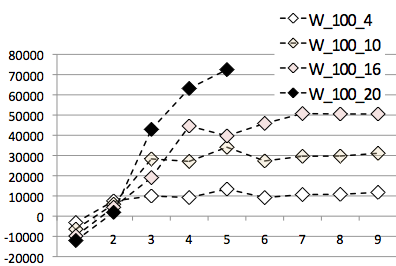}
\\ \small IRL,W,$|N|=100,\small ACD \in \{4,10,16,20\}$
\\ \small \textit{X axis}: Epoch number, \textit{Y axis}: Total payoff \\
\end{tabular}
\caption{Experimental results.}
  \vspace{-15pt}
%\ref{exps}
\end{figure}

\begin{center}
\begin{figure}
  \vspace{-20pt}
\begin{tabular}{ r | l }
\includegraphics[width=.40\textwidth]{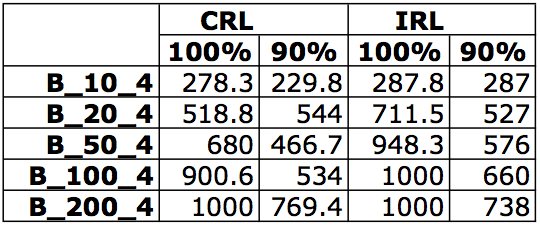} 
 &   
\includegraphics[width=.40\textwidth]{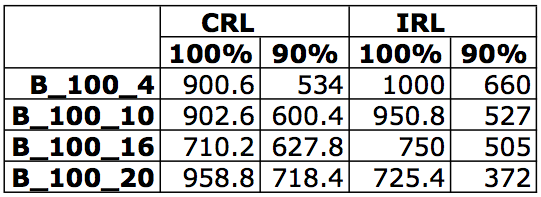} \\
\end{tabular}
\caption{Average convergence round per case.}
\vspace{-13pt}
%%\ref{exps}
\end{figure}
\end{center}
\section{Related work}
Early approaches towards learning norms either involve two agents iteratively playing a stage game towards reaching a preferred equilibrium, or models where the reward of each individual agent depends on the joint action of $all$ the other agents in the population. Other approaches consider that agents learn by iteratively interacting with a single opponent from the population \cite{SenIJCAI}, also considering the distance between agents \cite{MukherjeeSA08}. %Thus, in these approaches there are two agents learning to play a game at each time step. 
In contrast to this, in \cite{Yu2013}  the communication between agents is physically constrained and agents interact and learn with all their neighbors.
In these works agents learn rules of the road by playing a single role at each time step. We rather consider more realistic cases where agents do not share knowledge of their environment, they play multiple roles and interact with all their neighbors who also play multiple and maybe incompatible roles simultaneously. Finally, agents have role-specific  preferences on their strategies. 

Concerning the learning methods that have been used, Shoham and Tennenholtz  \cite{Shoham1997} proposed a reinforcement learning approach using the Highest Cumulative Reward rule. However this rule depends on the memory size of agents, as far as the history of agents' past strategy choices is concerned. The effects of memory and history of agents' past actions have also  been considered in the work reported by Villatoro et al \cite{Villatoro2011}, \cite{Villatoro2009},. Sen et al   \cite{SenIJCAI} studied the effectiveness of reinforcement methods also considering the influence of the population size, of the possible actions, the existence of different types of learners in the population, as well as the underlying network topology of agents \cite{SenCOIN}. In \cite{Yu2013} authors have proposed a learning method where each agent, at each time step interacts with all its neighbors simultaneously and use ensemble learning methods to compute a final strategy. In all experiments reported in the above mentioned studies Q-learners are more effective than other learners.   
In this work we propose two social learning Q-learning methods, according to which agents interact with all of their neighbors, considering all the roles in their social context. Agents compute role-specific strategies, while for a single role the decisions taken depends on the  feedback received from others, the existing constraints and role-specific preferences.  %We show the effectiveness of these social learning methods to compute society norms, by varying the population size and the connectivity of agents. 
\section{Conclusions and further work}
This article proposes two social, distributed reinforcement learning methods for agents to compute conventions concerning the use of common resources to perform joint tasks.  This happens via reaching agreements in their social context and in conjunction to computing semantic agreements on the representation of resources. To a greater extent than other models, agents play multiple roles in their social contexts, even with incompatible requirements and preferences. Results show that the proposed methods are efficient, despite the complexity of the problem considered. Further work concerns investigating (a) the effectiveness of hierarchical reinforcement learning techniques \cite{Barto} for computing hierarchical policies (for correspondences, scheduling strategies and joined tasks); (b) the tolerance of the methods to different payoffs of performing joined tasks, as well as to different exploration-exploitation schemes, and (c) societies with different types of learners.
%\textit{$\backslash$section*\{ACKNOWLEDGEMENTS\}}
\\
%\vspace
\\
%\textbf{About this article:} This article has been submitted with slightly different content and title to IJCAI 2015 and to another workshop. IJCAI 2015 allows this for workshops with no formal proceedings.

\bibliographystyle{splncs}%apalike}
{\small
\bibliography{norms_emergence}}

\begin{thebibliography}{10}

\bibitem{Eipstein}
Epstein, J.:
\newblock Learning to be thoughtless: Social norms and individual computation.
\newblock Computational Economics \textbf{18}(1) (2001)  9--24

\bibitem{NormLearning}
Savarimuthu, B.T.R.:
\newblock Norm learning in multi-agent societies.
\newblock In: Inf. Sc. Disc. Papers Ser. No. 2011/05,
 http://hdl.handle.net/10523/1690.
\newblock (2011)

\bibitem{MukherjeeSA08}
Mukherjee, P., Sen, S., Airiau, S.:
\newblock Norm emergence under constrained interactions in diverse societies.
\newblock In Padgham, L., Parkes, D.C., MŸller, J.P., Parsons, S., eds.: AAMAS
  (2), IFAAMAS (2008)  779--786

\bibitem{SenIJCAI}
Sen, S., Airiau, S.:
\newblock Emergence of norms through social learning.
\newblock In: Proceedings of  IJCAI'07, San Francisco, CA, USA, Morgan Kaufmann
  Publishers Inc. (2007)  1507--1512

\bibitem{Yu2013}
Yu, C., Zhang, M., Ren, F., Luo, X.:
\newblock Emergence of social norms through collective learning in networked
  agent societies.
\newblock In: Proceedings of AAMAS '13, (2013)  475--482

\bibitem{Bowling02}
Bowling, M., Veloso, M.:
\newblock Multiagent learning using a variable learning rate.
\newblock Artificial Intelligence \textbf{136} (2002)  215--250

\bibitem{Fudenberg}
Fudenberg, D., Levine, D.:
\newblock The theory in learning in games.
\newblock The MIT Press (1998)

\bibitem{Shoham1997}
Shoham, Y., Tennenholtz, M.:
\newblock On the emergence of social conventions: Modeling, analysis, and
  simulations.
\newblock Artif. Intell. \textbf{94}(1-2) (July 1997)  139--166

\bibitem{Vouros2013}
Vouros, G.:
\newblock Decentralized semantic coordination via belief propagation.
\newblock In: Proceedings of  AAMAS '13, (2013)  1207--1208

\bibitem{PutermanMDP}
Puterman, M.L.:
\newblock Markov Decision Processes: Discrete Stochastic Dynamic Programming.
  1st edn.
\newblock John Wiley \& Sons, Inc., New York, NY, USA (1994)

\bibitem{GuestrinPhD}
Guestrin, C.E.:
\newblock Planning Under Uncertainty in Complex Structured Environments.
\newblock PhD thesis, Stanford, CA, USA (2003) AAI3104233.

\bibitem{GuestrinCRL}
Guestrin, C.G., Lagoudakis, M., Parr, R.:
\newblock Coordinated reinforcement learning.
\newblock In: In Proceedings of ICML-2002 (2002)  227--234

\bibitem{Kok2006}
Kok, J.R., Vlassis, N.:
\newblock Collaborative multiagent reinforcement learning by payoff
  propagation.
\newblock J. Mach. Learn. Res. \textbf{7} (December 2006)  1789--1828

\bibitem{WattsStrogatz}
Watts, D.J., Strogatz, S.H.:
\newblock Collective dynamics of /`small-world/' networks.
\newblock Nature \textbf{393}(6684) (06 1998)  440--442

\bibitem{Barabasi}
Albert, R., l\'{a}szl— Barab\'{a}si, A.:
\newblock Statistical mechanics of complex networks.
\newblock Rev. Mod. Phys (2002)

\bibitem{Villatoro2011}
Villatoro, D., Sabater-Mir, J., Sen, S.:
\newblock Social instruments for robust convention emergence.
\newblock In: Proceedings of  IJCAI'11, AAAI Press (2011)
  420--425

\bibitem{Villatoro2009}
Villatoro, D., Sen, S., Sabater-Mir, J.:
\newblock Topology and memory effect on convention emergence.
\newblock In: Proceedings of 
  WI-IAT '09, Washington, DC, USA, IEEE Computer Society (2009)  233--240

\bibitem{SenCOIN}
Sen, O., Sen, S.:
\newblock Effects of social network topology and options on norm emergence.
\newblock In Padget, et al, eds.: Coordination, Organizations, Institutions
  and Norms in Agent Systems V. Volume 6069 of LNCS
\newblock Springer (2010)  211--222

\bibitem{Barto}
Barto, A.G., Mahadevan, S.:
\newblock Recent advances in hierarchical reinforcement learning.
\newblock Discrete Event Dynamic Systems \textbf{13}(1-2) (January 2003)
  41--77

\end{thebibliography}

\end{document}